\documentclass[10pt,a4paper,twoside]{article} 
\usepackage[top=2.1cm, bottom=2.6cm, inner=1.5cm, outer=1.5cm]{geometry}
\usepackage{graphicx}
\usepackage{titling}
\usepackage{hyperref}

\title{Structural and optical properties of monocrystalline and polycrystalline gold plasmonic nanorods}
\author{Luk\'a\v{s} Kej\'ik$^{*,1}$, Michal Hor\'ak$^{*,1}$, Tom\'a\v{s} \v{S}ikola$^{1,2}$, Vlastimil K\v{r}\'apek$^{1,2}$}

\predate{}
\postdate{}
\date{}

\begin{document}
\maketitle

$^{1}$ CEITEC, Brno University of Technology, Purky\v{n}ova 123, 612 00 Brno, Czech Republic

$^{2}$ Institute of Physical Engineering, Brno University of Technology, Technick\'a 2, 616 69 Brno, Czech Republic

$^{*}$ These authors contributed equally.

$^{*}$ lukas.kejik(at)ceitec.vutbr.cz

$^{*}$ michal.horak2(at)ceitec.vutbr.cz

\section*{Abstract}

Plasmonic structures are often fabricated by lithographic patterning of a thin metallic film. Properties of the thin film are intimately related to the quality of the resulting structures. Here we compare two kinds of thin gold films on silicon nitride membrane: 
a conventional polycrystalline thin film deposited by magnetron sputtering, and monocrystalline gold microplates which were chemically synthesised directly on the membrane's surface for the first time. Both pristine metals were used to fabricate plasmonic nanorods using focused ion beam lithography. Structural and optical properties of the nanorods were characterized by analytical transmission electron microscopy.
  
The dimensions of the nanorods in both substrates reproduced well the designed size of $240 \times 80\ \mathrm{nm^2}$ with the deviations up to 20\,nm in both length and width. The shape reproducibility was considerably improved among monocrystalline nanorods fabricated from the same microplate. Interestingly, monocrystal nanorods featured inclined boundaries while the boundaries of the polycrystal nanorods were upright.
Q factors and peak loss probabilities of the modes in both structures are within the experimental uncertainty identical. We conclude that the optical response of the plasmonic antennas is not deteriorated when the polycrystalline metal is used instead of the monocrystalline metal.

\section{Introduction}

Plasmonic antennas are metallic particles widely studied for their ability to control, enhance, and concentrate electromagnetic field~\cite{Novotny2011}. Strikingly, the field in the vicinity of plasmonic antennas, the so-called near field, can be focused into a deeply subwavelength region \cite{Schuller2010}. At the same time, the field is strongly enhanced with respect to the driving field, which can be e.g.~a plane wave. Focusing of the field stems from the excitation of localized surface plasmons (LSP) -- quantized oscillations of the free electron gas in the metal coupled to the evanescent electromagnetic wave propagating along the boundary of the metal. Thanks to this light-concentrating ability, plasmonic antennas have found applications in energy harvesting \cite{FeiGuo2014}, construction of metasurfaces \cite{Babocky2017, Babocky2018}, luminescence enhancement \cite{Butun2015}, or biodetection \cite{Salazar2018}.
 
Plasmonic antennas are conventionally prepared using electron beam lithography, which utilizes a mask fabricated in electron-sensitive polymer (resist), or more straightforwardly using direct milling by focused ion beam (FIB) \cite{Horak2018}. The energy of antenna's plasmon resonance is highly dependent on the choice of antenna material and its quality, surrounding material, as well as antenna size and shape \cite{Kelly2003}. 

Typical plasmonic antennas fabricated from metallic layers deposited by sputtering or evaporation techniques are polycrystalline by nature with randomly oriented grains of varying sizes \cite{Kobayashi2013, Hiramatsu2016, Parajuli2018}. The fabrication resolution and overall shape of the antenna are then significantly affected by the size of grains present in the antenna \cite{Huang2010} making each antenna of slightly different shape. The deviations in shape or the orientation of the antennas might then be detrimental e.g. for the performance of metasurfaces as they introduce random noise into their phase response. 
Although by optimizing the deposition parameters, the resulting grain size can be somewhat tuned \cite{Mahmoodi2017}, the presence of grain boundaries within an antenna is related to the lower resonance quality factor and higher relaxation rates \cite{Huang2007, Chen2010}.
In combination with other approaches, like template stripping, the layer quality might be improved even further \cite{McPeak2015}, and increased grain size can then allow antenna fabrication from a single grain. 
This can be achieved readily by chemical synthesis where each particle can be made as single crystal grain with great shape variability \cite{Langille2012}, but their precise placement on the substrate is hard to achieve. 
In an ideal case the whole set of antennas is fabricated from a single grain, possibly large metallic 2D monocrystal \cite{Hu2015,Chen2018} placed on a supporting substrate, alleviating the troubles with the size and shape reproducibility in the follow-up lithographic process. In the case of gold, there are several approaches to chemically reduce typically HAuCl$_{4}$ into large monocrystalline microplates in pathways based on aniline \cite{Guo2006}, two-component ionic liquids \cite{Kawasaki2007}, ethylene glycol \cite{Wu2015, Krauss2018, Kaltenecker2020} or tetraoctylammonium bromide (TOABr) \cite{Radha2010, Radha2011, Radha2012} where the latter two are the most prevalent. Although the improvement in optical properties of monocrystalline versus polycrystalline gold is not that radical in bulk \cite{Olmon2012}, it has been shown as a useful platform with improved propagation length of surface plasmons \cite{Pramassing2020} even in complex devices \cite{See2017, Frank2017, Boroviks2019, Siampour2020}.
Additionally, in terms of fabrication of antennas or more complex structures, monocrystalline gold allows fabrication with higher resolution, generally better defined shape and consequently better optical performance \cite{Hoffmann2016, Mejard2017}.

A detailed study, though, comparing the polycrystalline and monocrystalline antennas in terms of material properties and correlating the antenna shape with plasmon modes with high spatial and energy resolution, has been missing in the literature.

Here we show direct comparison of gold plasmonic nanorods fabricated either from monocrystalline or polycrystalline gold substrates using FIB lithography. We selected the nanorod shape due to well-isolated modes and an existence of analytical description \cite{Kalousek2012}. We characterize the material properties of the input substrates as well as the resulting nanorods, correlate the nanorod shape with its plasmonic properties and visualize the supported plasmonic modes by scanning transmission electron microscopy (STEM) in combination with electron energy loss spectroscopy (EELS) with nanometer spatial resolution not achievable by means of other methods.

\section{Methods}

\subsection{Fabrication of nanorod antennas}
Membranes with monocrystalline gold were prepared using the modified procedure by Radha and Kulkarni \cite{Radha2011}. Briefly, 4\,ml of TOABr solution in toluene (50\,mM) was stirred vigorously while 1.6\,ml of aqueous HAuCl$_{4}$ (25\,mM) was injected. The resulting mixture of two immiscible phases, aqueous and toluene-based, was stirred for approximately 2\,minutes. Aqueous component then became colourless while toluene component became red-coloured which is related to the phase transfer of (AuCl$_{4}$)$^{-}$ ions from aqueous solution into toluene using TOABr as a phase transfer agent. Toluene solution containing (AuCl$_{4}$)$^{-}$--\,TOABr precursor was then extracted and 1--2\,$\mu$l were typically applied onto SiN$_{x}$ TEM membrane, which was kapton-taped to a piece of silicon wafer for easier manipulation, on a preheated hot plate (75$^\circ$C). When toluene evaporated, the temperature was set to 140 or 145$^\circ$C held for 35--45\,hours. Such procedure resulted in monocrystalline gold microplates of various lateral sizes (tens of micrometers) and thickness (tens to hundreds of nanometers) and the remaining precursor residues were washed away in toluene, ethanol, and water baths, respectively. Note that the thinnest microplates (below 100\,nm) were pre-selected using bright field optical microscopy due to their partial transparency resulting in pink shade as opposed to thicker ones which reflected the light completely (see Fig.~\ref{figS1}). The final information about the thickness was obtained using EELS.

Polycrystalline gold layer (thickness of 30\,nm) was deposited by magnetron sputtering using Leica coater EM ACE600 directly (i.e.~with no adhesion layer) on a 30-nm-thick SiN$_{x}$ membrane (Agar Scientific).
 
Gold nanorods on both samples were fabricated by FIB lithography in dual beam FIB/SEM microscope FEI Helios using gallium ions with the energy of 30\,keV and ion beam current of 1.3\,pA. Note that the energy (the highest available) and the current (the lowest available) are optimized for the best spatial resolution of the milling \cite{Horak2018}. The nanorods were designed as rectangles with the length of 240\,nm and the width of 80\,nm. They were situated in the middle of a $1.5 \times 1$\,$\mu$m$^2$ metal-free rectangle, which is perfectly sufficient to prevent their interaction with the surrounding metallic layer. Note that the amount of residual gallium ions is negligible and does not influence the localized surface plasmon resonances \cite{Horak2018}. Results of the energy-dispersive X-ray spectroscopy (EDS) analysis are shown in Fig.~\ref{figS2}.

Finally, we remark that we have experienced issues with the mechanical stability of the membranes with monocrystalline nanorods which proved to be more brittle than untreated ones. 
In total we have fabricated five monocrystalline samples containing microplates of suitable thickness (below 100\,nm). Three samples have been destroyed during the fabrication process or sample manipulation and the last two were characterized in the transmission electron microscope (TEM). For the latter ones it was possible to obtain majority of the properties, although no membrane survived complete analysis. Therefore, a complete characterization of a single sample turned out impossible and we used different samples instead. We have not experienced any stability issues for the polycrystalline substrates. We suppose that the low stability of the monocrystalline substrates is related to the membrane heating, its loading with the growth solution or by the following washing procedure in various liquid solvents. Conversely, a continuous thin film of polycrystalline gold does not undergo any wet treatment and can even contribute to the mechanical strength of the membrane unlike the finite-size monocrystalline microplates with a rather low coverage. 

\subsection{Analytical methods}
Surface roughness of the resulting polycrystalline and monocrystalline gold substrates was analysed by atomic force microscopy (AFM) using Bruker Dimension Icon in tapping ScanAsyst mode.

Analytical transmission electron microscopy characterization including STEM-EELS was performed using TEM FEI Titan equipped with Super-X spectrometer for EDS and GIF Quantum for EELS operated at the primary electron energy of 300\,keV.
Prior to STEM-EELS experiments all samples were cleaned in oxygen/argon plasma for 10\,seconds to prevent the carbon contamination \cite{Horak2018}.
Beam current was set around 0.2\,nA and the full-width at half-maximum (FWHM) of the zero-loss peak (ZLP) was in the range from 0.1\,eV to 0.15\,eV. 
We set the convergence semi-angle to 10\,mrad and the collection semi-angle to 20.5\,mrad. We note that these parameters are not critical for the signal-to-background ratio of the setup \cite{Horak2020}. The dispersion of the spectrometer was 0.01\,eV/pixel. We recorded spectrum images with the pixel size of 2\,nm. Every pixel consists of 1 EEL spectrum whose acquisition time was adjusted to use the maximal intensity range of CCD camera in the spectrometer and avoid its overexposure. EEL spectra were integrated over rectangular areas consisting of tens of pixels in the areas corresponding to maximum loss probability of studied LSP resonances. Obtained spectra were divided by the integral intensity of the ZLP (the energy window for integration was set as $-1$\,eV to $+1$\,eV) to transform measured counts to a quantity proportional to the loss probability. We fitted the peaks in the EEL spectra by Gaussian function to determine the peak energy and its spectral width (i.e., FWHM). EEL maps were calculated by dividing the map of integrated intensity at the plasmon peak energy with the energy window of 0.1\,eV by the map of the integral intensity of the ZLP. 

Simulations of EELS spectra have been carried out using boundary element method (BEM)~\cite{GarciadeAbajo2002} with a software package MNPBEM~\cite{Waxenegger2015}. The dielectric function of gold was taken from Ref.~\cite{Johnson1972} and the dielectric constant of the silicon nitride membrane was set equal to 4, which is a reasonable approximation in the considered spectral region~\cite{Schmidt2014morph} taking into account non-stoichiometric composition of the membrane~\cite{Horak2018}. The polycrystalline nanorods have been represented by a rectangular prism and the monocrystalline nanorods by a truncated pyramid with the top face reduced by 30~nm from each side in comparison to the bottom face (i.e., with the length and width smaller by 60 nm). The edges of the nanorods have been rounded with a radius of 10 nm.

\section{Structural properties of the nanorods}

\begin{figure}[ht!]
  \begin{center}
    \includegraphics[width=16cm]{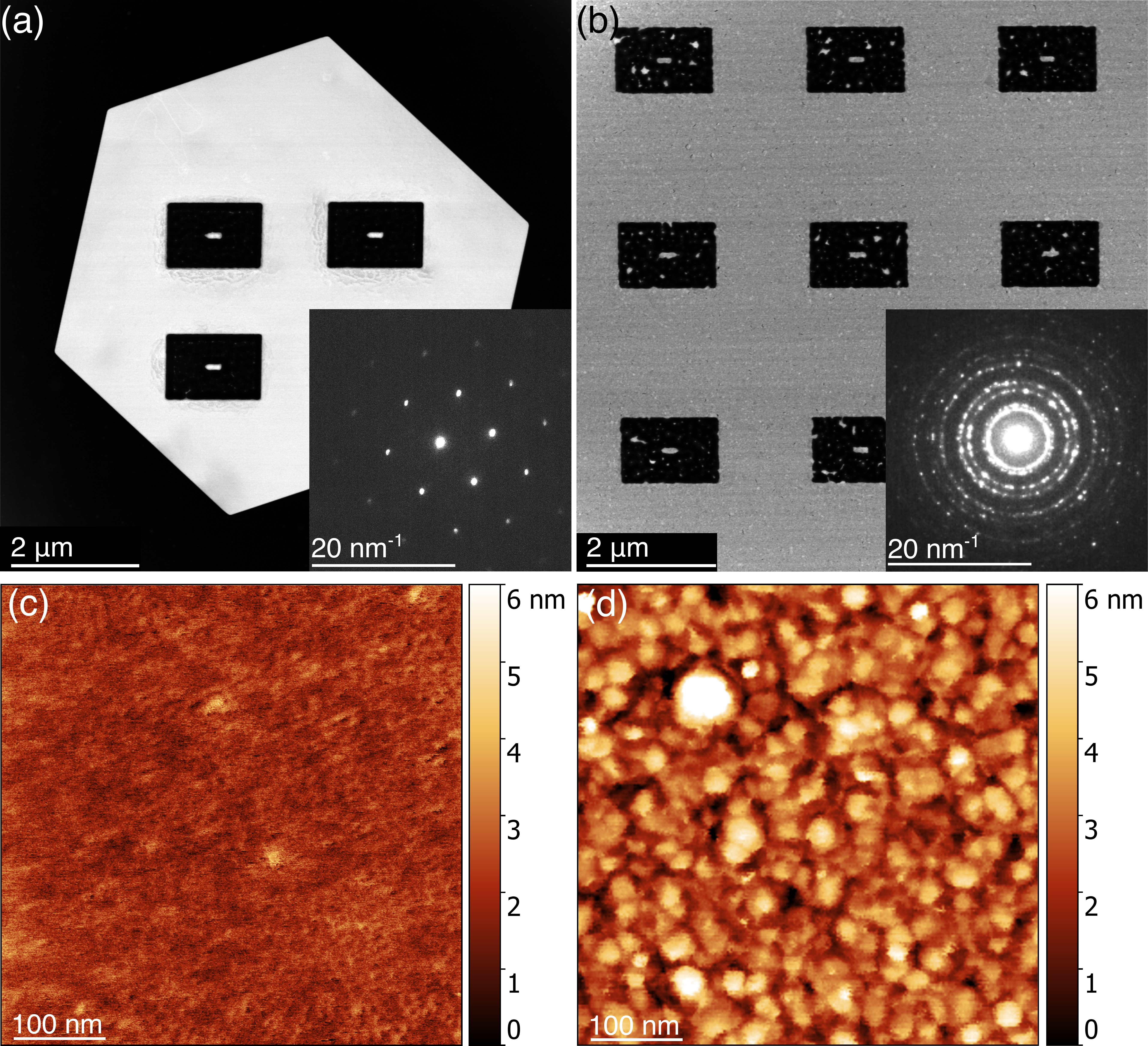}
    \caption{\label{figNO1} (a,b) STEM ADF image of (a) a monocrystalline microplate and (b) a polycrystalline thin film after the fabrication of nanorods. Insets show electron diffraction patterns recorded at the area of the film far from the nanorods. (c,d) AFM images of the surface of a typical (c) monocrystalline microplate and (d) a polycrystalline thin film.}
  \end{center}
\end{figure}

Thin monocrystalline microplates and polycrystalline films of gold have been characterized by selected-area electron diffraction (SAED) in TEM and by AFM. 
Fig.~\ref{figNO1}(a) shows a monocrystalline microplate with fabricated nanorods and corresponding electron diffraction pattern recorded at an area away from the nanorods. 
The diffraction pattern exhibits a 6-fold rotational symmetry and corresponds to the face-centered cubic (FCC) Au crystal lattice viewed along the [111] direction. A point-like symmetric character of the pattern indicates the single-crystalline nature of the synthesized Au microplates laying flat on the membrane with (111)-oriented basal plane.

\begin{figure}[ht!]
  \begin{center}
    \includegraphics[width=16cm]{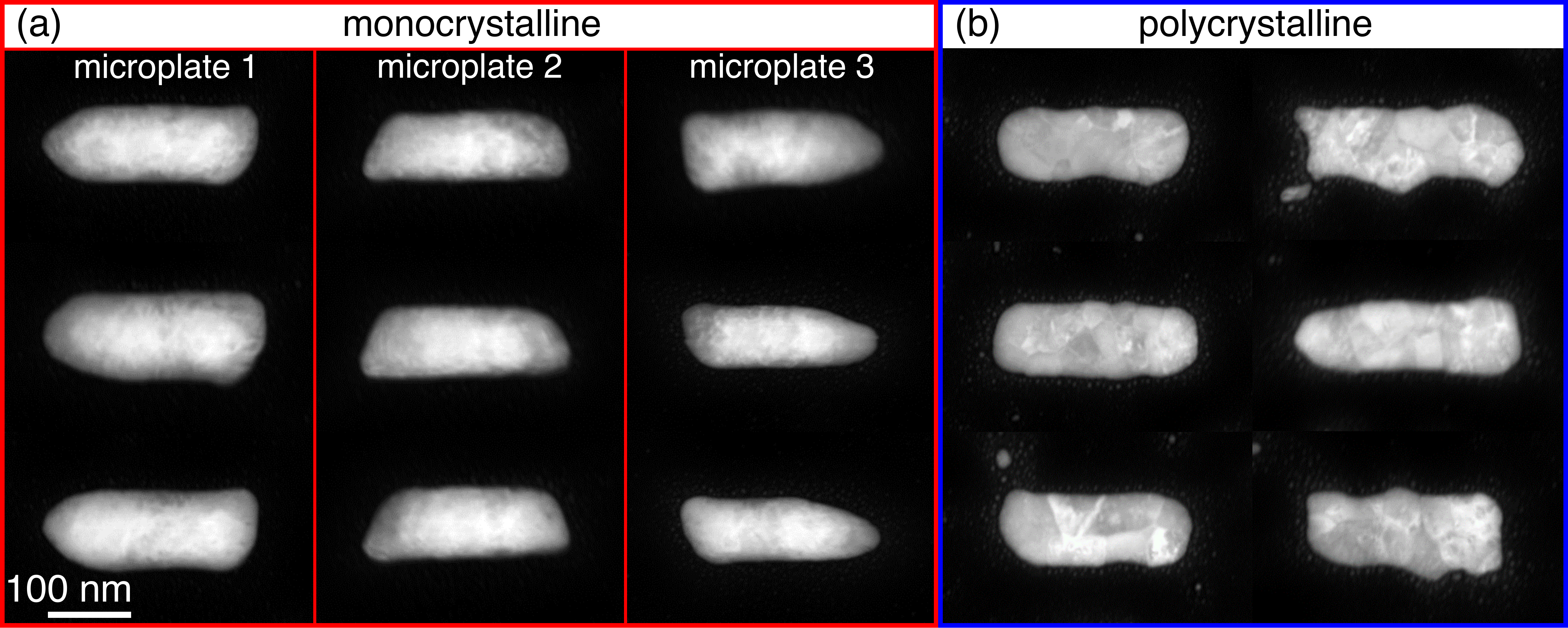}
    \caption{\label{figNO2}STEM ADF images of gold nanorods fabricated from (a) monocrystalline and (b) polycrystalline gold substrates. Note that monocrystalline nanorods are divided into columns according to the microplate from which they were fabricated.}
  \end{center}
\end{figure}

Similar characterization of a polycrystalline thin film is shown in Figure~\ref{figNO1}(b).
For the selected scale, the individual grains of the polycrystal are not well visible [see Fig.~\ref{figNO2}(b) for a zoomed-in images]. Nevertheless, a circular diffraction pattern clearly corresponds to a polycrystalline sample with random orientation of individual grains. 

AFM has been performed for a different set of pristine samples fabricated under the same growth conditions as the samples with nanorods. It reveals rather flat surface of both monocrystalline [Fig.~\ref{figNO1}(c)] and polycrystalline [Fig.~\ref{figNO1}(d)] samples, with root mean square surface roughness on average around 0.6\,nm for the monocrystalline and 1.2\,nm for the polycrystalline substrates. Larger roughness of the polycrystalline film is related to a stochastic nature of the gold sputtering.

The overview of resulting nanorod shapes which underwent EELS analysis is shown in Fig.~\ref{figNO2}. Monocrystalline nanorods are rather regular, with smooth boundaries. However, the tips of the rods are faceted and deviate from the rectangular shape. The area surrounding the nanorods is clean and smooth (i.e., it shows no features in the STEM). Polycrystalline nanorods are of more irregular shape, with coarse boundaries and overall larger deviations from the desired rectangular shape. The irregularity and larger variance are related to the random orientation of grains in the polycrystalline metal layer, for which different sputtering rates apply during the FIB lithography. Individual grains of gold are visible in the STEM images, with typical sizes between 20 and 50\,nm. Nevertheless, polycrystalline nanorods lack the faceted tips and some may reproduce the desired rectangular shape better than monocrystalline nanorods [e.g. the first and the second rod in Fig.~\ref{figNO2}(b)]. The area around the polycrystalline nanorods is covered by gold crystallites due to incomplete removal of the gold layer again related to different removal rate of randomly orientated gold grains. We note that it is not possible to remove these crystallites simply by increasing the removal time as it would result into thinning and finally a destruction of the membrane underneath. 
Upon close inspection, monocrystalline nanorods fabricated from the same gold monocrystal exhibit very similar overall shape with the tips faceted in the same manner which is also reflected in much smaller deviations in terms of e.g.~antenna length (Tab.~\ref{Tab1}). Across the different microplates though the facets look very different. 
When comparing the respective orientation of gold monocrystal and the resulting nanorods, some of the monocrystal's side facets seem to be imprinted into the antenna shape (see~Fig.~\ref{figS3}). Therefore, the fabrication results might be further improved by a suitable alignment of the monocrystal side facets with the FIB milling direction.  

\begin{figure}[ht!]
  \begin{center}
    \includegraphics[width=16cm]{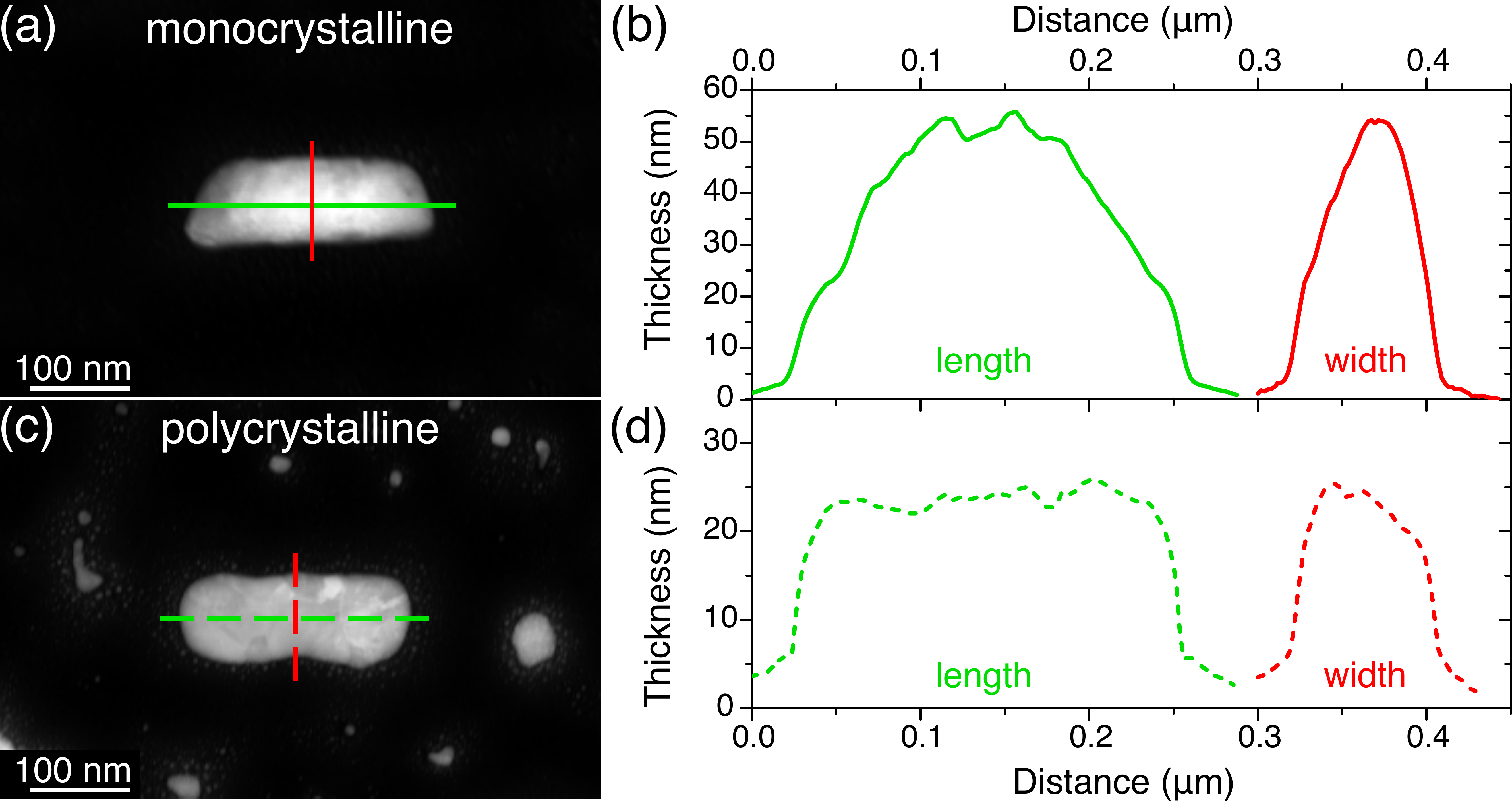}
    \caption{\label{figNO3} (a,c) STEM ADF images of representative plasmonic nanorods fabricated from (a) a monocrystalline microplate and (c) a polycrystalline thin film with cross-sections corresponding to (b,d) thickness profiles along the long and short axes of (b) the monocrystalline and (d) polycrystalline nanorod determined by EELS. The color of thickness profile curve corresponds to the color of the axis indicated in (a,c).}
  \end{center}
\end{figure}

Representative nanorods with the targeted length and width $240 \times 80$\,nm$^2$ are shown in Fig.~\ref{figNO3}(a,c) with the corresponding thickness profiles along their length and width in Fig.~\ref{figNO3}(b,d).
We have used EELS to retrieve the thickness of the nanorods using the same approach described in Ref.~\cite{Horak2018} based on Refs.~\cite{egerton, Mitchell2006, Iakoubovski2008}. Polycrystalline nanorods exhibit upright boundaries with a steep change in the thickness from zero to maximum over a distance of about 10\,nm related to the diameter of the ion beam whose nominal FWHM is around 3\,nm. On the other hand, monocrystalline nanorods have strongly inclined boundaries mostly due to microplate's anisotropy with possible gold redeposition. 

The dimensions of all the antennas have been determined as the maximum extent of the white contrast in the STEM annular dark field (ADF) images along the longitudinal (length $L$) and transverse (width $W$) direction, and the height $H$ has been determined as the average height of the central part of the nanorod. The dimensions of the monocrystalline nanorods were $L=(232 \pm 9)$\,nm, $W=(82 \pm 7)$\,nm, and $H=(52 \pm 5)$\,nm, while the dimensions of the polycrystalline nanorods were $L=(238 \pm 14)$\,nm, $W=(84 \pm 3)$\,nm, and $H=(30 \pm 3)$\,nm (see Tab.~\ref{Tab1} for a complete summary). The targeted lateral dimensions have been reproduced well. The height of the monocrystalline microplates cannot be quantitatively controlled and due to low mechanical stability of the monocrystalline samples we measured the height only after the fabrication of the antennas together with the EELS characterization of plasmon modes. We could fabricate the polycrystalline nanorods (for which the height is controlled rather accurately) after the characterization of the monocrystalline antennas, assuring the same thickness. However, this would necessarily result into some deviation in the alignment and adjustment of the electron microscopes used for the fabrication of the nanorods and their characterization, restricting the comparability of both types of the nanorods. Therefore, we preferred fabrication and characterization of both samples without any delay and need for realignment of the microscopes at the price of different nanorod heights.
In addition, the inclined boundaries of the monocrystalline nanorods resulted in the LSP energies nearly the same as for the polycrystalline nanorods, which is optimum for the comparison of both systems despite their different height.

\section{Optical properties of the nanorods}

\begin{figure*}[ht!]
  \begin{center}
    \includegraphics[width=17cm]{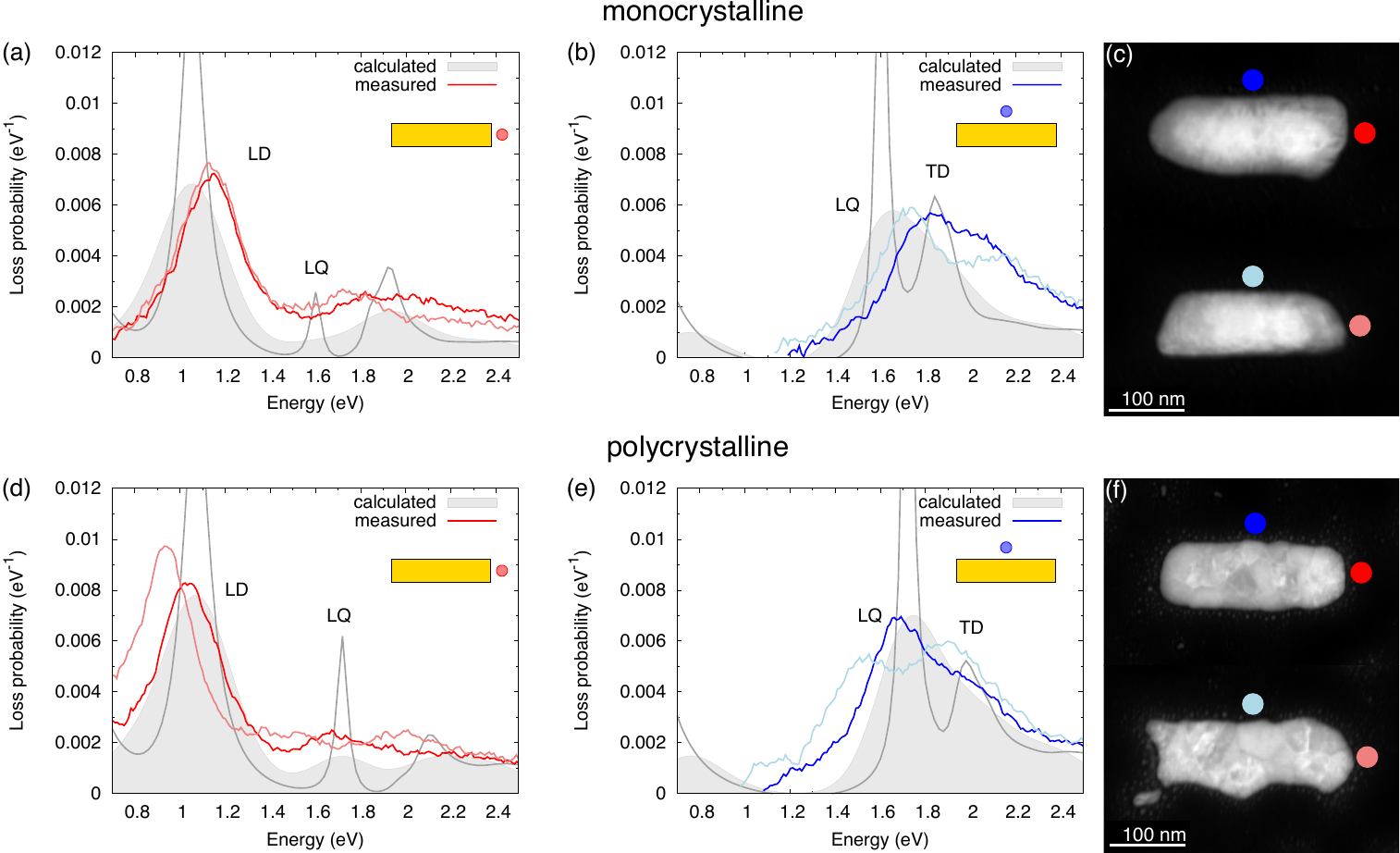}
    \caption{\label{figNO4} Loss probability spectra for (a,b) monocrystalline and (d,e) polycrystalline nanorods recorded for the electron beam located at (a,d) the center of the short edge and (b,e) the center of the long edge for two different (c) monocrystalline and (f) polycrystalline nanorods. Experimental spectra are displayed by red or blue lines. Calculated spectra are displayed by gray lines, calculated spectra broadened by a Gaussian with 0.3\,eV FWHM to reproduce the instrumental broadening are displayed by gray shaded area. Most significant LSP modes are labelled as LD (longitudinal dipole), LQ (longitudinal quadrupole), and TD (transverse dipole) mode. (c,f) STEM ADF images of the selected nanorods with color spots indicating the position of the electron beam with colors corresponding to the respective EEL spectra.}
  \end{center}
\end{figure*}

\begin{figure}[ht!]
  \begin{center}
    \includegraphics[width=12cm]{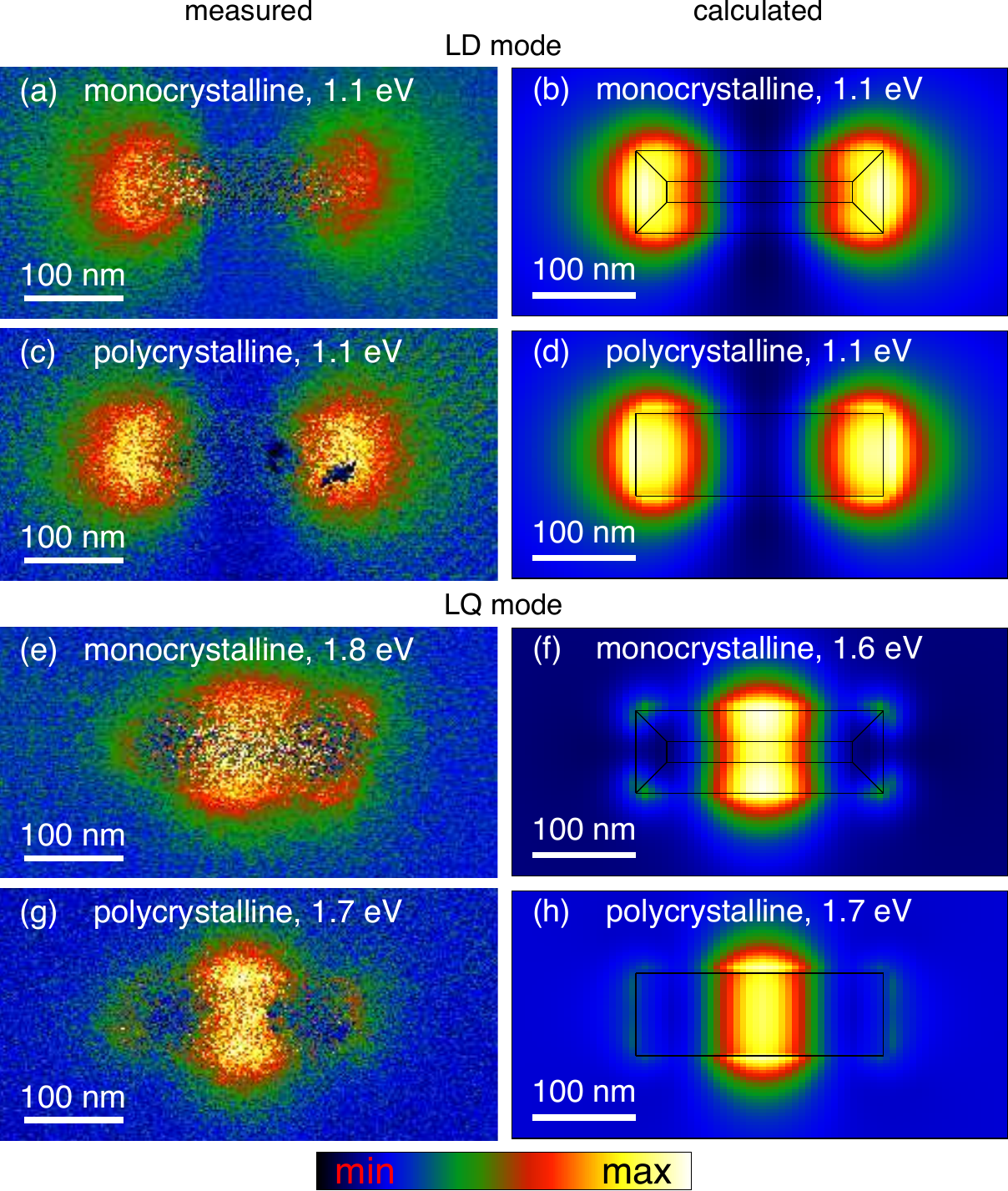}
    \caption{\label{figNO5} Spatial distribution of the loss probability at the energy of (a-d) the LD mode and (e-h) the LQ mode. Left panels (a,c,e,g) show experimental maps for nanorods displayed in upper part of Fig.~\ref{figNO4}(c,f). Right panels (b,d,f,h) show calculated values with the contour of the nanorod displayed by the black lines. The energy of the mode is indicated at the top part of each panel. The color scale at the bottom is common to all panels.}
  \end{center}
\end{figure}

Two red and blue curves in Fig.~\ref{figNO4} represent typical experimental loss probability spectra of two monocrystalline and two polycrystalline nanorods. The lowest-order LSP mode is observed around 1.1\,eV and by inspecting energy-filtered maps of the loss intensity in Fig. \ref{figNO5}(a,c), it is identified as the longitudinal dipole (LD) mode. 
The peak at 1.8\,eV is attributed to two spectrally overlapping modes, the longitudinal quadrupole (LQ) mode and the transverse dipole (TD) mode.
It is difficult to resolve the LQ and TD modes in the experiment, as well as the higher-order modes. Nevertheless, the energy-filtered maps of the loss intensity for the peak energy [Fig. \ref{figNO5}(b,d)] suggest LQ mode to be dominant at the peak energy. 

Both main peaks of experimental spectra corresponding to monocrystalline antennas are slightly blue-shifted with respect to polycrystalline ones. Since the differences in dielectric function of monocrystalline and polycrystalline gold have been previously found as insignificant~\cite{Olmon2012}, we attribute the blue shift to the differences in the shape and thickness of the nanorods. 
The experimental loss probability spectra are accompanied by those calculated using MNPBEM shown as grey lines. The simulations take into account actual height profiles of the antennas as determined by EELS, i.e., the height of 50\,nm with inclined boundaries for the monocrystalline nanorods and the height of 30\,nm with upright boundaries for the polycrystalline nanorods. Note that the larger thickness causes a shift towards higher mode energies while the presence of inclined boundaries causes shift to lower energies, as shown in Fig.~\ref{figS4}.
To account for the experimental spectral broadening of our EELS system, the simulated spectra were convoluted with a Gaussian of 0.3\,eV FWHM shown as shaded gray spectra. 
These broadened spectra are then in good agreement with the experimental ones. 

By fitting the spectral profile of the modes by a Gaussian, we obtained the LSP resonance energy $E$ and Q factor defined as the LSP resonance energy divided by FWHM for both main modes. Mean energies of the LD mode in experiment averaged to $(1.14 \pm 0.03)~\mathrm{eV}$ for the monocrystalline nanorods and $(1.02 \pm 0.05)~\mathrm{eV}$ for the polycrystalline nanorods. 
Taking into account the experimental resolution of EELS, estimated as 0.1\,eV for our setup, no significant difference in the energy of LD mode has been observed between monocrystalline and polycrystalline nanorods.
The simulations accounting for the dispersion in length and width of nanorods [10~\% deviations from the mean value are assumed, i.e., $L=(240 \pm 24)~\mathrm{nm}$, $W=(80 \pm 8)~\mathrm{nm}$] yielded the energies of the LD mode in the range $(1.06 \pm 0.08)~\mathrm{eV}$ for monocrystalline and $(1.08 \pm 0.07)~\mathrm{eV}$ for polycrystalline nanorods, in good agreement with the experimental values taking into account the experimental error and uncertainty of the shape and dimensions of model nanorods.

To confirm the nature of the LSP modes induced in gold nanorods, we extracted EEL maps to visualize the modes at the peak energies. Fig.~\ref{figNO5}(a,c) shows EEL maps at the 1.1\,eV energy for monocrystalline and polycrystalline gold nanorods with a loss probability distribution typical for the LD mode. Similarly, the EEL maps at the higher energy peak at 1.8\,eV for monocrystalline and 1.7\,eV for polycrystalline nanorods [Fig.~\ref{figNO5}(e,g)] imply that the longitudinal quadrupole mode is more prevalent than the transverse dipole mode at the peak energy. EEL maps for both modes show only minute differences between monocrystalline and polycrystalline nanorod, although the modes in polycrystalline one can be resolved a bit easier. For detailed interpretation of the EEL maps we refer to Ref.~\cite{Krapek2020}.
The experimental EEL maps were complemented by the simulated EEL maps with a very good agreement in both LD [Fig.~\ref{figNO5}(b,d)] and LQ [Fig.~\ref{figNO5}(f,h)] mode after taking into account the differences in the nanorod morphology. 

Table~\ref{Tab1} summarizes morphological as well as optical properties of all studied antennas. In total, we have analysed 7 polycrystalline and 9 monocrystalline nanorods which were fabricated using in total three monocrystalline microplates. In terms of the nanorods' size the average length of the polycrystalline nanorods was closer to the desired length (240\,nm) though with larger deviations, the width of the monocrystalline nanorods was closer to the desired width (80\,nm) also with larger deviations. The average energy of LD mode for the monocrystalline nanorods is slightly larger and with smaller overall dispersion than for polycrystalline ones, the difference is even more pronounced in the case of LQ mode. Similarly, extracted Q factors are slightly larger in case of monocrystalline nanorods for both modes. Overall the average parameters of the nanorods' plasmon resonance for monocrystalline and polycrystalline nanorods are quite similar. Although when we focus on the nanorods fabricated from within the same gold microplate, the dispersion of measured parameters is significantly reduced, especially in terms of LD and LQ mode energy.

\begin{table*}[h!]
\caption{\label{Tab1} Summary of average dimensions and plasmon resonance parameters of polycrystalline and monocrystalline gold nanorods characterized by EELS followed by parameters averaged only over nanorods originating from the same gold monocrystal. Note that the numbers in parentheses in the left column correspond to the number or analysed nanorods.}

\footnotesize{
\begin{tabular}{l|c|c|c|c|c|c}

                                      & \textbf{Length (nm)} & \textbf{Width (nm)} & \textbf{LD energy (eV)} & \textbf{LD Q factor} & \textbf{LQ energy (eV)} & \textbf{LQ Q factor} \\ \hline
\textbf{Polycrystalline - all (7)}              & $238 \pm 14$         & $84 \pm 3$          & $1.01 \pm 0.05$         & $3.2 \pm 0.4$        & $1.67 \pm 0.08$         & $4.5 \pm 0.6$        \\ \hline
\textbf{Monocrystalline - all (9)}        & $232 \pm 9$          & $82 \pm 7$          & $1.14 \pm 0.03$       & $3.4 \pm 0.3$      & $1.78 \pm 0.03$         & $5.1 \pm 0.4$        \\ \hline
\multicolumn{1}{r|}{\textbf{microplate 1 (3)}} & $243 \pm 3$          & $90 \pm 5$          & $1.12 \pm 0.03$       & $3.1 \pm 0.2$      & $1.81 \pm 0.01$        & $4.9 \pm 0.4$        \\ \hline
\multicolumn{1}{r|}{\textbf{microplate 2 (3)}} & $231 \pm 2$      & $80 \pm 2$      & $1.13 \pm 0.01$       & $3.4 \pm 0.1$      & $1.75 \pm 0.02$        & $5.0 \pm 0.4$        \\ \hline
\multicolumn{1}{r|}{\textbf{microplate 3 (3)}} & $223 \pm 2$      & $76 \pm 5$          & $1.17 \pm 0.02$       & $3.7 \pm 0.2$      & $1.77 \pm 0.01$        & $5.3 \pm 0.4$       
\\ \hline
\textbf{Theory polycrystalline}        & $240 \pm 24$          & $80 \pm 8$          & $1.08 \pm 0.07$       & $2.8 \pm 0.2$      & $1.88 \pm 0.08$         & $5.5 \pm 0.1$       
\\ \hline
\textbf{Theory monocrystalline}        & $240 \pm 24$          & $80 \pm 8$          & $1.06 \pm 0.08$       & $2.9 \pm 0.2$      & $1.62 \pm 0.07$         & $4.4 \pm 0.5$       

\end{tabular}
}
\end{table*}

\section{Conclusions}
We have prepared monocrystalline and polycrystalline plasmonic nanorods and compared their structural and optical properties. Most pronounced differences have been identified regarding the vertical interfaces of nanorods, inclined for the monocrystalline nanorods but upright for polycrystalline ones. Monocrystalline nanorods outperform the polycrystalline ones in terms of slightly reduced size and shape fluctuations. They also benefit from clean metal-free parts, while for polycrystalline nanorods these parts contain metal crystallites due to incomplete metal removal. On the other hand, monocrystalline nanorods systematically deviate from the desired shape due to preferable termination with specific crystal planes. The membranes with monocrystalline nanorods also suffered from increased brittleness.

As for the optical properties, we have observed no significant differences for both crystallinities. The plasmon resonance energies, Q factors, and loss probability magnitudes detected by EELS show no difference up to experimental accuracy. The optical response of nanorods is not deteriorated when the polycrystalline metal replaces the monocrystalline metal. To conclude, polycrystalline plasmonic antennas represent a fully-fledged alternative of monocrystalline antennas.

\section*{Acknowledgements}
We acknowledge the Czech Science Foundation (grant No.~19-06621S), and Ministry of Education, Youth and Sports CR (projects CzechNanoLab Research Infrastructure, No.~LM2018110 and CEITEC 2020, No.~LQ1601).
M. H. acknowledges the support of Thermo Fisher Scientific and CSMS scholarship 2019.

\newpage
\section*{Supporting information}

\renewcommand{\thefigure}{S\arabic{figure}}
\setcounter{figure}{0} 

\begin{figure*}[h!]
  \begin{center}
    \includegraphics[width=160mm]{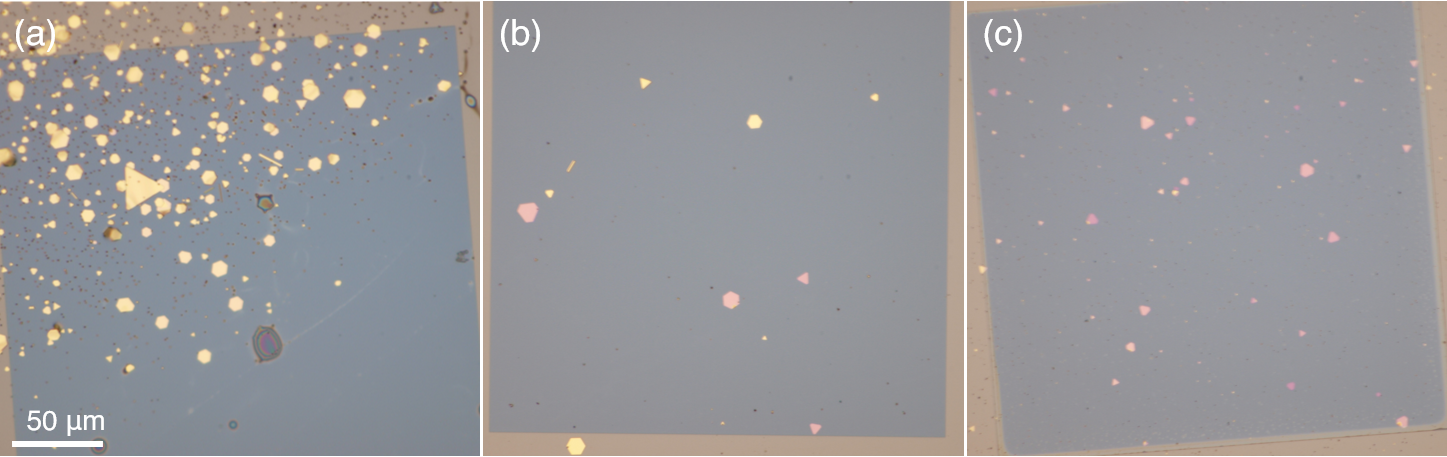}
    \caption{\label{figS1} (a,b,c) Bright field optical images of gold microplates on 30\,nm SiN$_x$ membrane (square $250 \times 250$\,$\mu$m). Membrane containing (a) thick microplates with more than 100\,nm thickness which are completely opaque for the incident light, (b) combination of completely opaque thicker and partially transparent thinner microplates which have slight pink shade, (c) thin microplates with the thickness well below 100\,nm - the thinner the microplate the darker the shade of pink color.}
  \end{center}
\end{figure*}

\begin{figure*}[h!]
  \begin{center}
    \includegraphics[width=160mm]{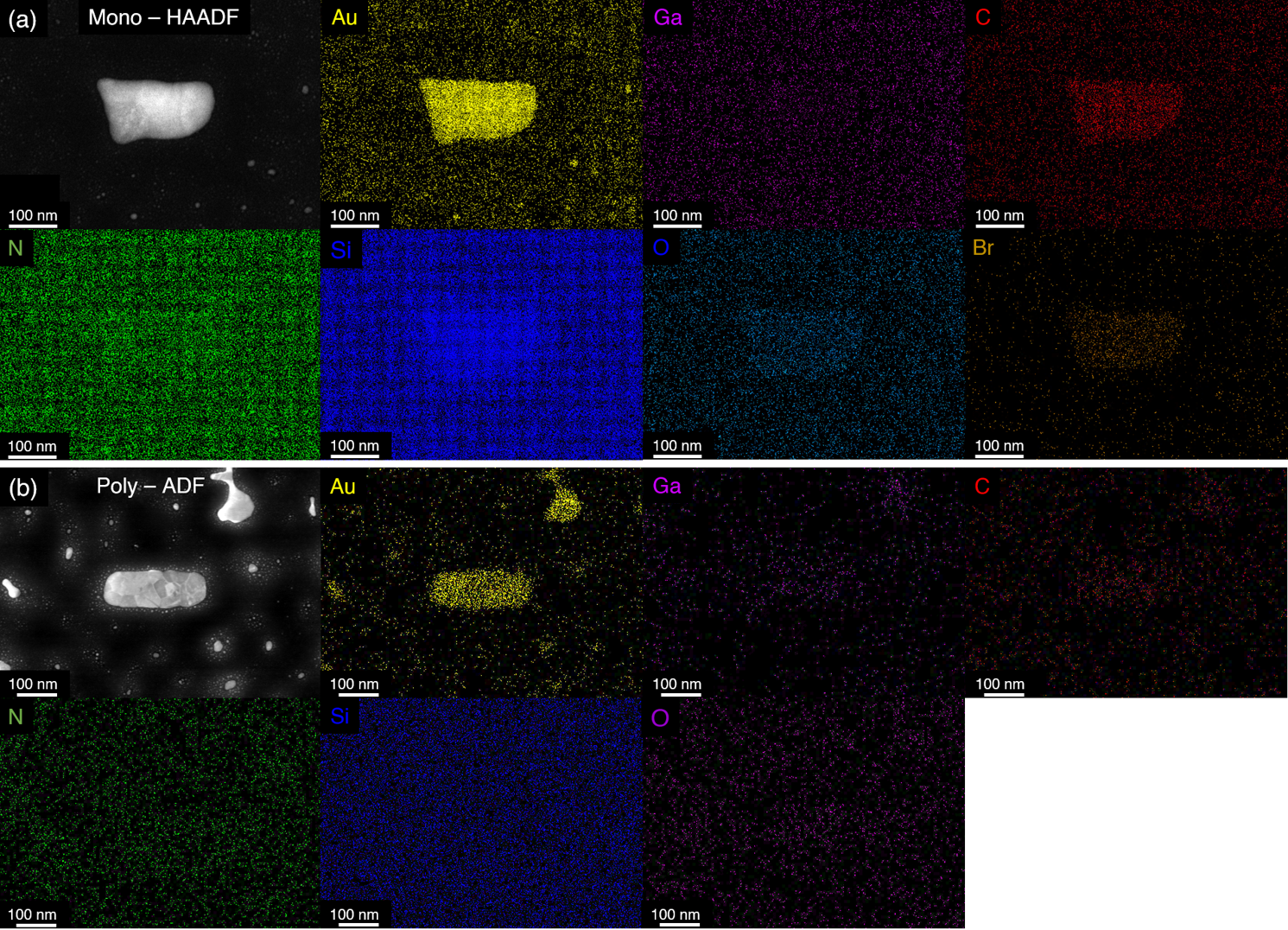}
    \caption{\label{figS2} (a,b) Energy-dispersive X-ray spectroscopy (EDS) analysis of fabricated (a) monocrystalline and (b) polycrystalline gold nanorods with the maps of all relevant elements.}
  \end{center}
\end{figure*}

\begin{figure*}[h!]
  \begin{center}
    \includegraphics[width=160mm]{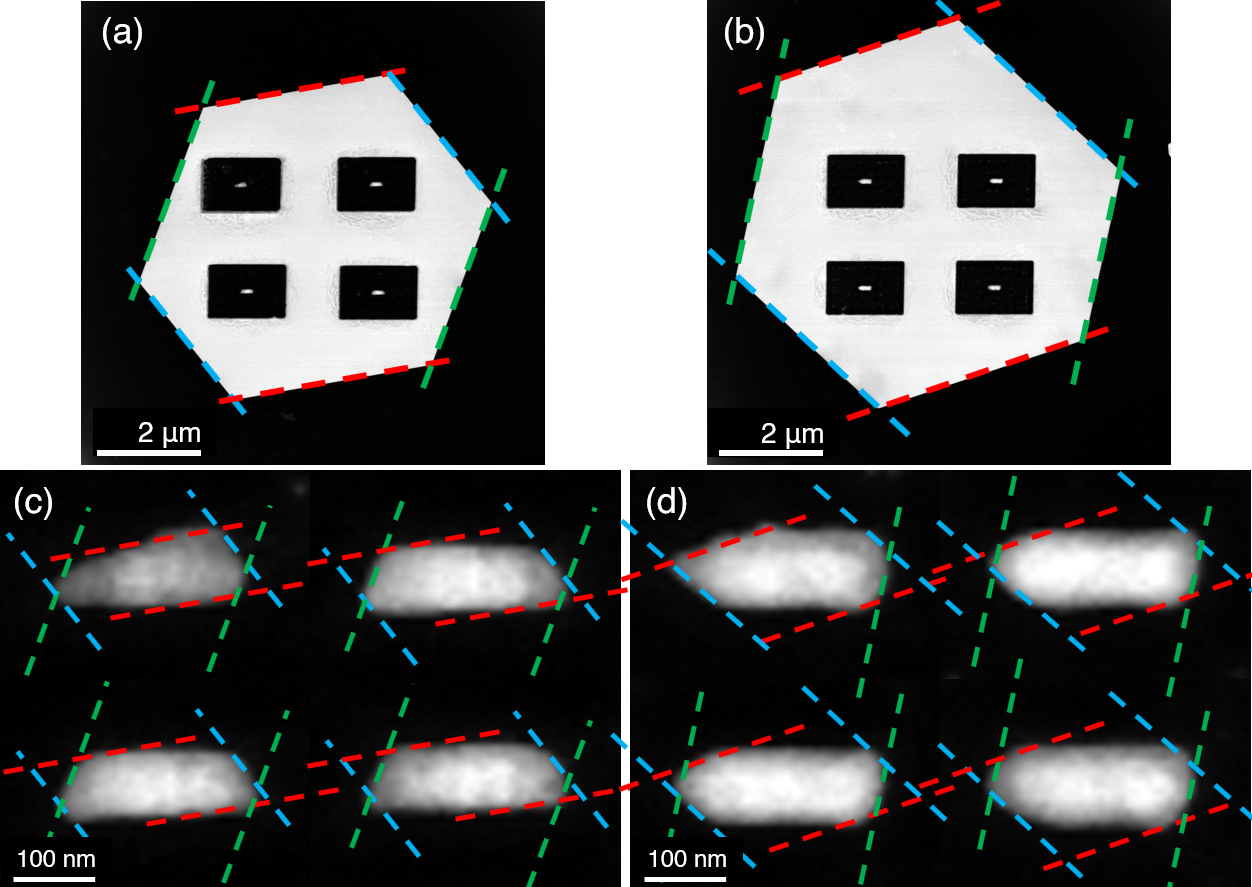}
    \caption{\label{figS3} (a,b) ADF STEM images of selected gold microplates after FIB fabrication of nanorod antennas. (c,d) Details of fabricated monocrystalline nanorods with the respective "mother" microplate's facets highlighted: nanorods in (c) correspond to the microplate in (a) and nanorods in (d) correspond to the microplate in (b).}
  \end{center}
\end{figure*}

\begin{figure*}[h!]
  \begin{center}
    \includegraphics[width=160mm]{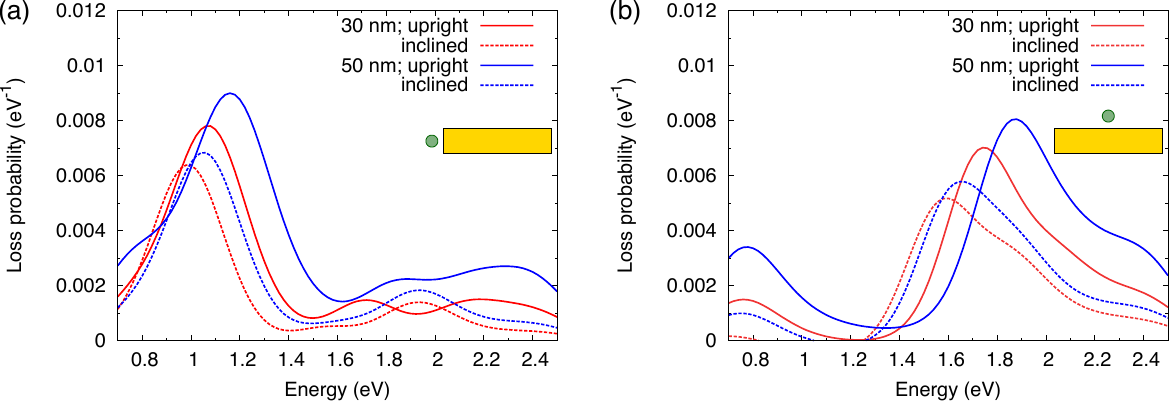}
    \caption{\label{figS4} (a,b) Calculated EEL spectra for gold nanorods of 30\,nm thickness (red curves) and 50\,nm thickness (blue curves) with upright boundaries (solid lines) and inclined boundaries (dashed lines) for electron beam positioned at the (a) shorter and (b) longer nanorod's edge. Note that all the spectra were broadened by a Gaussian with 0.3\,eV FWHM to reproduce the experimental broadening.}
  \end{center}
\end{figure*}

\end{document}